\documentclass[reprint,nofootinbib,amsmath,amssymb,aps,floatfix,superscriptaddress,twocolumn]{revtex4-2}
\usepackage{graphicx}
\usepackage{dcolumn}
\usepackage[colorlinks,linkcolor=magenta,anchorcolor=cyan,citecolor=blue]{hyperref}
\usepackage{bm}
\usepackage[multiple]{footmisc}
\usepackage{lipsum}
\usepackage{soul}

\begin{document}
\title{Observational Features of Thin Accretion Disk Around Rotating Regular Black Hole}

\author{Zhen Li} 
\email{zhen.li@just.edu.cn}
\affiliation{School of Science, Jiangsu University of Science and Technology, Zhenjiang 212100, China}

\date{\today}

\begin{abstract}
Rotating regular black hole, as a promising extension beyond general relativity, offer a phenomenological model that resolves spacetime singularities. In this study, we investigate the observational features of thin accretion disk around a well-known rotating regular black hole, which introduce an exponential converge factor $e^{-k/r}$ to the black hole mass M, where $k$ is the regular parameter. By studying the effects of the regular parameter on key quantities such as energy, angular momentum, and the innermost stable circular orbits of a test particle, we are able to analyze the radiative flux, temperature, and differential luminosity of the thin accretion disk in this rotation regular black hole spacetime. By using the ray-tracing method, we also obtain the bolometric images of thin accretion disk around this rotating regular black hole, with various black hole spins and inclination angels. Our results show that the regular parameter significantly impacts the observables, enhancing the radiation efficiency of thin accretion disk and contracting the lensing bands of black hole image, compared to Kerr black hole. These effects become more pronounced for rapidly rotating regular black hole and large inclination angels, making them more detectable in astrophysical observations.
\end{abstract}
\maketitle

\section{Introduction}

General relativity (GR) as the best theory of gravitational interaction is strongly supported by recent observations of gravitational waves \cite{gw1,gw2,gw3} and black hole images \cite{bh1,bh2}. However, it faces the spacetime singularity problem, which is widely considered non-physical and highlights the limitations of general relativity \cite{sg1,sg2}. Alternative black hole without singularity or the so-called regular black hole is a promising way to explore the physics beyond GR. There are many kinds of regular black hole models \cite{rr0, rr1, rr2}, including the famous Bardeen \cite{bar} and Hayward models \cite{hay}. Recently, a new regular black hole model attracts the attention of gravitational community, which modifies the black hole mass with an exponential converge factor \cite{re1,re2} (see also a geodesically complete version \cite{re3}),  yielding regular spacetime metrics. The rotating counterparts of this regular black hole are also obtained in \cite{rre1,rre2,rre3}. Interestingly, this black hole model has a asymptotic flat spacetime metric in its core, just like the peace in the eye of storm \cite{rre2}. While some studies have examined their phenomenology, focusing on black hole images, quasinormal modes, energy extraction mechanism, primordial black holes, and observational tests \cite{pre1,pre2,pre3,pre4, pre40, pre41, pre42, pre5,pre6,pre7,pre8,pre9,pre10,pre11,pre12}, broader astrophysical tests remain limited and warrant further investigation.

In recent years, advancements in astronomical techniques have significantly enhanced the precision on the accretion disks observations \cite{ao1,ao2,ao3,ao4}, allowing for detailed measurements of observable quantities such as the temperature and luminosity. Moreover, with the Event Horizon Telescope, we have achieved the unprecedented goal of capturing images of black holes \cite{bh1,bh2}, which essentially depict the accretion disks surrounding them \cite{lk1,lk2}. These developments provide crucial empirical evidence for testing black hole models beyond GR. Therefore, studying the impact of regular black hole models on the accretion disk is not only significant but also is feasible in comparing with observations. It could pave the way for addressing key questions about regular black holes and their behavior, offering insights into fundamental physics in strong gravitational fields.

The accretion disk is usually modeled as thin accretion disk, which was originally proposed by Novikov-Thorne \cite{t1}, and Page–Thorne \cite{t2}, as well as Shakura and Sunyaev \cite{t3}, assuming that the accretion disk is geometrically thin and optically thick. Under these assumptions, the accretion disk radiates approximately as a blackbody at each radius \cite{t1,t2,t3}. This allows for simplified yet highly effective descriptions of disk behavior, making it a widely adopted model for studying accretion processes in various astrophysical scenarios, including black holes. Even recently, there are plenty of works on studying the observational signatures of thin accretion disk in different black hole spacetimes \cite{ts1,ts2,ts3,ts4,ts5,ts6,ts7,ts8,ts9,ts10,ts11,ts12}, focusing on the effects on the flux, the temperature, luminosity, images, and so on.

There are some related works on imaging this regular black hole based on empirical intensity profile which may not consist with the accretion disk model. In addition, they are mainly focusing on the image of non-rotating ones \cite{pre41,pre11}, or only the shadow of rotating ones \cite{pre4,pre40}. There are no related works on the image of thin accretion disk around rotating regular black hole. However, the rotating one is more realistic than a static one, since the astronomical body always has angular momentum. The researches of its effects on thin accretion disk and black hole image are still lacking. In this study, we will investigate the observational features of thin accretion disk around rotating regular black hole. We will thoroughly study how the regular parameter changes the properties of thin accretion disk and its observables, as well as its effects on the black hole image (or more precisely the bolometric image), laid down the foundation for further study on testing alternatives regular black holes through the observation of accretion disk.

This paper is organized as follows: In Sec.~\ref{sec2}, we will introduce the rotating regular black hole and discuss some useful quantities or radii of studying the observational features for both time-like and null-like geodesics. In Sec.~\ref{sec3}, we will study the emission properties of thin accretion disk in this rotating regular spacetime, and showing how the regular parameter affects observables of accretion disk. Then, in Sec.~\ref{sec4}, we will obtain the realistic bolometric image of thin accretion disk based on the ray-tracing method, and comparing the results with Kerr black hole, discussing the effects of rotating regular black hole on the lensing bands and the images. In Sec.~\ref{sec5}, we will make a conclusion and discussion on the results of this work.

\section{Geodesic properties of rotating regular black hole}\label{sec2}

The metric of rotating regular black hole written in the Boyer-Lindquist coordinates is \cite{rre1,rre2,rre3}, 
\begin{equation}\label{metric}
d s^{2}=g_{t t} d t^{2}+g_{r r} d r^{2}+g_{\theta \theta} d \theta^{2}+g_{\phi \phi} d \phi^{2}+2 g_{t \phi} d t d \phi,
\end{equation}
where the metric components are given by 
\begin{align}\label{mc}
g_{t t}&= -\left(1-\frac{2 M r \mathrm{e}^{-k / r}}{\Sigma}\right)\quad   g_{t\phi}=-\frac{2 a M r \mathrm{e}^{-k / r}}{\Sigma} \sin ^{2} \theta \nonumber\\
g_{r r} &=\frac{\Sigma}{\Delta}\qquad \qquad \quad\quad\quad\quad\quad g_{\theta \theta} =\Sigma \nonumber\\
g_{\phi\phi}&=\left(r^{2}+a^{2}+\frac{2 M r a^{2} \mathrm{e}^{-k / r}}{\Sigma} \sin ^{2} \theta\right) \sin ^{2} \theta ,
\end{align}
with $\Sigma=r^{2}+a^{2} \cos ^{2} \theta$ and $\Delta=r^{2}+a^{2}-2 M re^{-k / r}$. $\Delta=0$ will result in two roots $r_-$ and $r_+$, which are respectively the inner and event horizon of this rotating regular black hole. The black hole mass $M$, spin $a$, and regular parameter $k$ are assumed to be positive. The Kerr metric is a special case of the metric (\ref{metric}) when $k=0$. Here and after, we use geometrized units with $G=c =1$.

Similar to Kerr black hole, the motion of particles or photons in this spacetime could be described with three constants of motion \cite{geo}. They are energy $E$, angular momentum along the spin axis $L$, and the Carter integral $Q$, and the last two quantity could also be replaced by energy rescaled quantities as $\lambda \equiv L/E$ and $\eta \equiv Q/E$. The geodesic equations for a test particle or photon, in the spacetime (\ref{metric}), will take forms as 
\begin{align}
\frac{\Sigma}{E} p^{r}&= \pm_{r} \sqrt{\mathcal{R}(r)}, \label{pr}\\
\frac{\Sigma}{E} p^{\theta} &= \pm_{\theta} \sqrt{\Theta(\theta)},  \label{p8}\\
\frac{\Sigma}{E} p^{\phi}&=\frac{a}{\Delta}\left(r^{2}+a^{2}-a \lambda\right)+\frac{\lambda}{\sin ^{2} \theta}-a , \label{pp}\\
\frac{\Sigma}{E} p^{t}&=\frac{r^{2}+a^{2}}{\Delta}\left(r^{2}+a^{2}-a \lambda\right)+a\left(\lambda-a \sin ^{2} \theta\right) ,\label{pt}
\end{align}
where $p^\nu\,(\nu=r,\theta,\phi,t)$  are the components of four momentum, $\pm_{r}$ and $\pm_{\theta}$ respectively represent the sign of $p^r$ and $p^\theta$, $\mathcal{R}(r)$ and $\Theta(\theta)$ are called the radial and angular potential defined as 
\begin{align}
\mathcal{R}(r)&=\left(r^{2}+a^{2}-a \lambda\right)^{2}-\Delta\left[\mu^2 r^2+\eta+(\lambda-a)^{2}\right], \\
\Theta(\theta)&=\eta+a^{2} \cos ^{2} \theta-\lambda^{2} \cot ^{2} \theta .
\end{align}
For photons $\mu=0$. While for massive particles, one is free to set $\mu=1$, which indicates the conserved quantities $E$, $\lambda$, and $\eta$ are all normalized with particle mass $\mu$. We are considering the  conserved quantities of time-like geodesics in the per unit mass basis. 

\subsection{Key quantities of time-like geodesics}
We assume that the particles composing the thin accretion disk are all massive and moving in circular orbits at the equatorial plane, which follow the time-like geodesics. In order to study the observational properties of thin accretion disk, we need first to know several key quantities of these particles. Although we are interested in the particles moving circularly in the equatorial plane where $Q=0$ and $\theta = \pi/2$, we still provide a very general form to the key quantities in the following. The general form is more concise and universal than the specific cases.

In general, the angular velocity of a test particles could be obtained through the geodesic equations (\ref{pr}-\ref{pt}) and expressed as a function of the metric components (\ref{mc}), they are 
\begin{equation}\label{omega}
\Omega=\frac{-\partial_{r} g_{t \phi} \pm \sqrt{\left(\partial_{r} g_{t \phi}\right)^{2}-\left(\partial_{r} g_{t t}\right)\left(\partial_{r} g_{\phi \phi}\right)}}{\partial_{r} g_{\phi \phi}}.
\end{equation}
With the expression of $\Omega$, one could obtain the energy and angular momentum per unit particle mass of a test particle as follows,
\begin{equation}\label{en}
E = -\frac{g_{t t} + g_{t \phi} \Omega}{\sqrt{-g_{t t} - 2 g_{t \phi} \Omega - g_{\phi \phi} \Omega^{2}}},
\end{equation}
\begin{equation}\label{an}
L = \frac{g_{t \phi} + g_{\phi \phi} \Omega}{\sqrt{-g_{t t} - 2 g_{t \phi} \Omega - g_{\phi \phi} \Omega^{2}}} ,
\end{equation}
where we set $\mu =1$. Another quantity we wish to know is the time component of four velocity of this test particle, which takes the following form
\begin{equation}\label{ut}
u^{t} = \frac{1}{\sqrt{-g_{t t}-2 \Omega g_{t \varphi}-\Omega^{2} g_{\varphi \varphi}}}.
\end{equation}
The last quantity we need to consider is the radius of innermost stable circular orbit (ISCO) denote as $r_{\text{ISCO}}$ of this test particle, which could computed by setting the second radial derivative of radial potential equals to zero,
\begin{equation}
\frac{d^2\mathcal{R}(r)}{dr^2}\bigg| _{r_{\text{ISCO}}} \,= 0 .
\end{equation}
Dose not like the Kerr metric, we do not have an analytical formula for $r_{\text{ISCO}}$, rather we could only compute it numerically. We adopted the numerical method from \cite{che}.

The $r_{\text{ISCO}}$ for the circular orbits in the equatorial plane with different $k$ and black hole spin $a$ are shown in Fig.~{\ref{rms}}. We can see that as the spin parameter $a$ increases, the radius of $r_{\text{ISCO}}$ decreases monotonically for all $k$ values. This is the same to the behavior predicted by Kerr black hole models, where higher spins lead to more compact stable orbits. For fixed values of $a$, an increase in $k$ leads to an overall decrease in $r_{\text{ISCO}}$. This indicates that the parameter $k$ introduces a deviation from the standard Kerr black hole, effectively decreasing the radius of stable orbits. The $r_{\text{ISCO}}$ is always greater than the event horizon $r_+$ as we expected. The $r_{\text{ISCO}}$ and event horizon will become coincident as the spin reaches a maximum value. When beyond the maximum spin, the rotating regular black hole will have no horizons. The larger regular parameter $k$, the smaller the maximum spin.

\subsection{Key radii of null-like geodesics}
For photons $\mu =0$, it will follow the null-like geodesics. Although it is not related to the composition of thin accretion disk, it is crucial to the ray-tracing method, which is used to obtain the black hole image. We will discuss several important radii that will be used a lot in the ray-tracing method and thus in the black hole imaging process.

The roots of radial potential for the rotating regular black hole classify the photons trajectories of traced rays into different cases, either falling into black hole, or scattering back to infinity. They will determine the behavior of photons and thus directly influence the imaging results of rotating regular black hole. The roots could be obtained by solving the following equation $\mathcal{R}(r)=0$ or equivalently,
\begin{equation}
r^{4}+\mathcal{A} r^{2}+\mathcal{B} r+\mathcal{C}=0,
\end{equation}
where
\begin{equation}
\begin{aligned}
\mathcal{A} & =a^{2}-\eta-\lambda^{2} ,\\
\mathcal{B} & =2 M e^{-k/r}\left[\eta+(\lambda-a)^{2}\right]>0 ,\\
\mathcal{C} & =-a^{2} \eta .
\end{aligned}
\end{equation}
We have written the above equation in a quadratic form in order to compare with the results in Kerr metric. However, it is not a quadratic equation anymore because here $B$ have a factor $e^{-k/r}$ and thus is $r$ dependent. Actually, they have more than four roots, and we could only solve it numerically. Luckily, when $k\ll1$, most of the extra roots are deep inside horizon and appears in complex conjugates, which allow us to focus on the largest four roots (compared in the real part of the roots) denoted as $r_1$, $r_2$, $r_3$, and $r_4$, and $\Re (r_1)< \Re (r_2)< \Re (r_3)< \Re (r_4)$. Then the classification of photon trajectories are similar to Kerr black hole \cite{lk1,lk2}.

\begin{figure}[htbp]
  \centering
    \includegraphics[scale = 0.52]{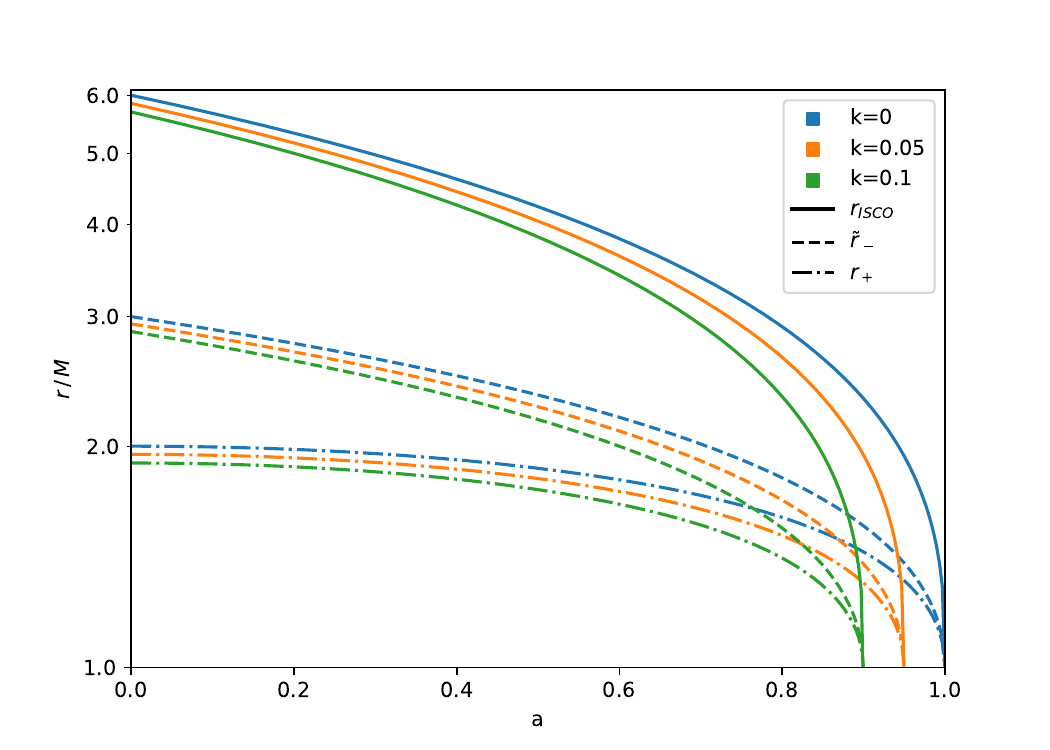}
\caption{The dependence of various radii on the spin parameter $a$ and the regular parameter $k = 0, 0.05, 0.1$. The plot shows the behavior of the innermost stable circular orbit radius $r_{\text{ISCO}}$, the inner photon shell radius $\tilde{r}_- $, and the event horizon radius $r_+$. Increasing $k$ leads to an inward shift in all these three characteristic radii.}
\label{rms}
\end{figure}

Next we wish to know the critical root $\tilde{r}_{\pm}$, which characterizing the boundary of photon shells or the bounded photon orbits in the rotating regular black hole spacetime. The radii of bounded photon orbits $\tilde{r}$ could be obtained by solving following equation,
\begin{equation}
\mathcal{R}(r)\bigg| _{\tilde{r}} =\frac{d\mathcal{R}(r)}{dr}\bigg| _{\tilde{r}}=0.
\end{equation}
For $\tilde{r}>r_+$, to fulfill the above equation, the conserved quantities $\lambda$, $\eta$ can only take the following unique values
\begin{align}
\tilde{\lambda}&=a+\frac{\tilde{r}}{a}\left[\tilde{r}-\frac{2 \tilde{\Delta}}{\tilde{r}-M-N\tilde{r}}\right], \label{cr1}\\
\tilde{\eta}&=\frac{\tilde{r}^{3}}{a^{2}}\left[\frac{4 \tilde{\Delta}(M -N\tilde{r})}{(\tilde{r}-M - N\tilde{r})^{2}}-\tilde{r}\right],\label{cr2}\\
N &= \frac{k}{\tilde{r}^2}e^{-k/\tilde{r}} ,\label{cr3}
\end{align}
where $\tilde{\Delta} = \tilde{r}^{2}+a^{2}-2 M \tilde{r}e^{-k / \tilde{r}}$. In the equatorial plane, we have $\tilde{\eta} = 0$, then this condition will provide us two radii that greater than the horizon $\tilde{r}_\pm >r_+$, $\tilde{r}_-$ and $\tilde{r}_+$ corresponding to the inner and out boundary radius of photon shell, respectively. All the radius of bounded orbits $\tilde{r}$ lie within the range $\tilde{r}\in[\tilde{r}_-,\,\tilde{r}_+]$. Bounded orbits are not stable and will form a critical curve in the observer's image plane, see Sec.~\ref{sec4}. We have shown the inner edge of photon shell $\tilde{r}_-$ for different $k$ and black hole spin $a$ in Fig.~{\ref{rms}}

We can see that as the spin parameter $a$ increases, $\tilde{r}_-$ decreases for all values of $k$. For a fixed $a$, the value of $\tilde{r}_-$ decreases slightly with increasing $k$. This indicates that the deviation parameter $k$ alters the black hole geometry, pushing the $\tilde{r}_-$ inward. The gap between $\tilde{r}_-$ and $r_+$ becomes narrower as $a$ approaches the maximum, indicating that high-spin black holes exhibit a more compact structure. The presence of $k$ speed up this narrowing, suggesting its role in suppressing the distinction between the two radii. As the spins approach their maximum, $\tilde{r}_-$ also coincides with the event horizon, the same to $r_{\text{ISCO}}$,  of the rotating regular black hole.

\section{Emission properties of thin accretion disks}\label{sec3}

\begin{figure}[htbp]
  \centering
    \includegraphics[scale = 0.52]{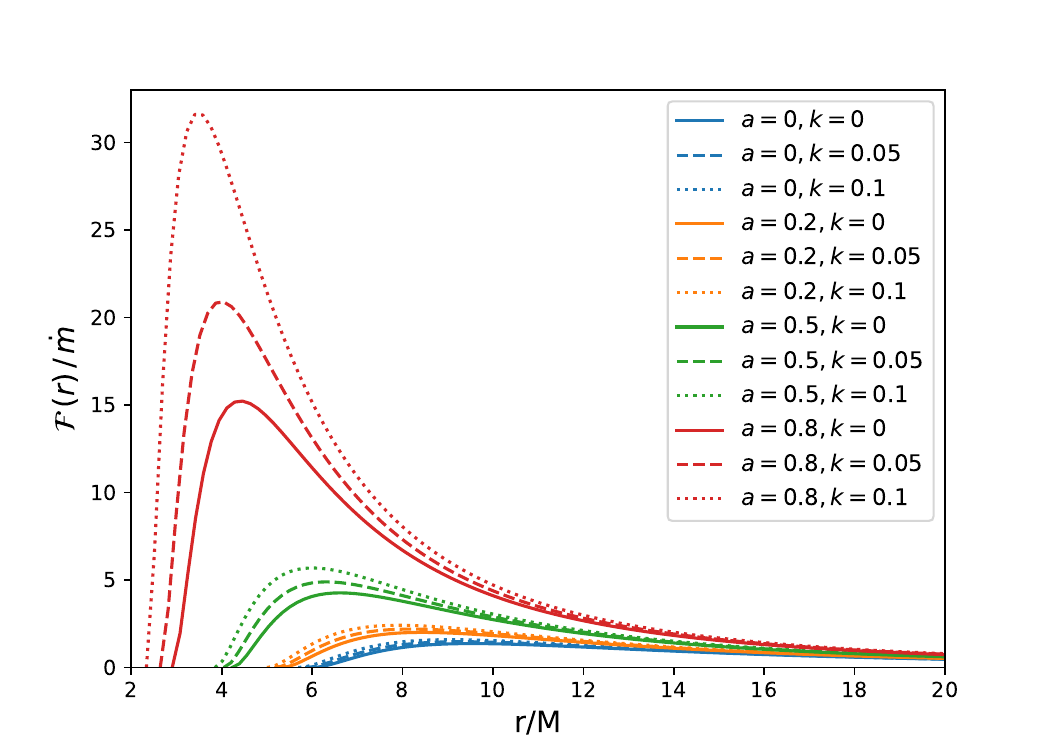}
\caption{The radiative flux per accretion rate of thin accretion disk $\mathcal{F}(r)/\dot m$ as a function of $r/M$ for different black hole spin $a$ and regular parameter $k$. The results of Kerr black hole ($k = 0$) are also provided as references. All the values are multiplied by $10^5$ for convenience.}
\label{f}
\end{figure}

In thin accretion disk model \cite{t1,t2,t3}, we have the radiative flux $\mathcal{F}(r)$ which represents the energy emitted per unit area and per unit time at radius $r$ from the thin accretion disk. It takes form as
\begin{equation}\label{flux}
\mathcal{F}(r)=-\frac{\dot m }{4 \pi \sqrt{-g}} \frac{\Omega_{, r}}{(E-\Omega L)^{2}} \int_{r_{I S C O}}^{r}(E-\Omega L) L_{, \tilde{r}} d \tilde{r},
\end{equation}
where $\dot m $ is the mass accretion rate, and here we consider $\dot m $ is much smaller than the Eddington accretion rate, where the thin accretion disk model is valid.
$g$ is the metric determinant of three dimensional subspace $(t,r,\phi)$ 
\begin{equation}
g = g_{rr}(g_{tt}g_{\phi\phi}-g_{t\phi}^2).
\end{equation}
$\Omega$, $E$, and $L$ are given by Eq.~(\ref{omega}), (\ref{en}) and (\ref{an}), respectively. $\Omega_{, r}$ and $L_{, r}$ represent the radial derivatives of angular velocity and angular momentum of a test particle. We only consider that the thin accretion disk lie within the equatorial plane where $\theta =\pi/2$, so the metric components as well as the expressions for $\Omega$, $E$, and $L$ will simplify a lot. We have shown the the dependence of radiative flux per accretion rate for various black hole spin $a$ and regular parameters $k$ in Fig.~{\ref{f}}. The values in this figure are all multiplied by $10^5$ for convenience.

We can see that for a non-rotating black hole ($a = 0$), the radiative flux corresponding to different $k$ exhibit minimal differences, indicating weak effects of the regular parameter $k$. However, as the spin parameter increases, the deviations between different $k$ grow significantly, especially the inner region of the thin accretion disk. The larger $k$, the lager radiative flux, which means greater radiation efficiency. The figure also shows the combined influence of $a$ and $k$ on the physical properties of the thin accretion disk. The inner radii where radiative flux approach to zero are the corresponding $r_{\text{ISCO}}$. We could see that they are shifted to smaller radius for large black hole spin and regular parameter. The values of radiative flux also increases as the spin and regular parameter increase. In the far region, they tend to be the same regardless the black hole spin $a$ and regular parameter $k$.

\begin{figure}[htbp]
  \centering
    \includegraphics[scale = 0.52]{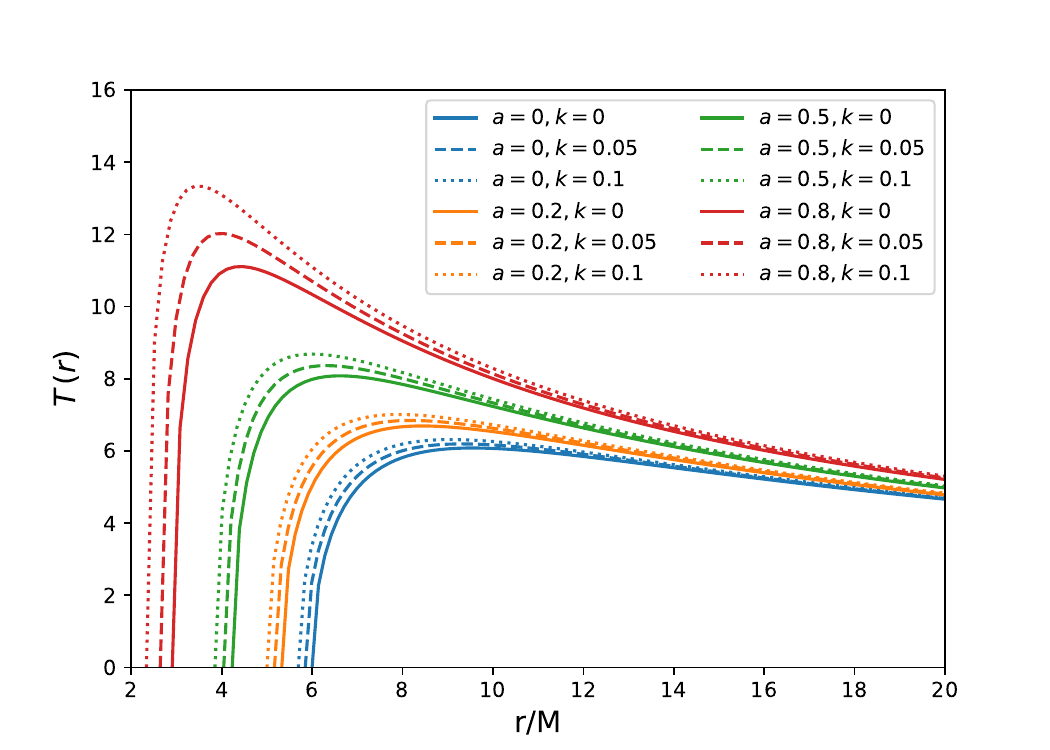}
\caption{The temperature of thin accretion disk $T(r)$ as a function of $r/M$ for different black hole spin $a$ and regular parameter $k$. We set the Stefan–Boltzmann constant $\sigma =1$, and the values in this figure are multiplied by $10^2$ for convenience}
\label{t}
\end{figure}

Assuming at each radius $r$, the thin accretion disk is in the state of thermal equilibrium, we could then obtain the temperature according to the Stefan-Boltzmann law
\begin{equation}
T = \sqrt[4]{\mathcal{F}(r)/\sigma},
\end{equation}
where $\sigma$ is the Stefan–Boltzmann constant. The distributions of temperature as a function of radial distance with different black hole spin $a$ and regular parameter $k$ are shown in Fig.~{\ref{t}}. The results of Kerr black hole ($k = 0$) are also provided as references. The Stefan–Boltzmann constant $\sigma$ is set to be 1, and the values in this figure are all multiplied by $10^2$ for convenience.

From Fig.~{\ref{t}}, we know that as the spin parameter increases, the temperature distributions show larger deviations, particularly at the inner region of the accretion disk. The larger black hole spin $a$ and regular parameter $k$, the higher temperature of the thin accretion disk. The inner radii where temperature approach to zero are also the corresponding $r_{\text{ISCO}}$. At larger radii, the temperature curves converge, reflecting the decreasing influence of both black hole spin $a$ and regular parameter $k$ in the far region.

\begin{figure}[htbp]
  \centering
    \includegraphics[scale = 0.52]{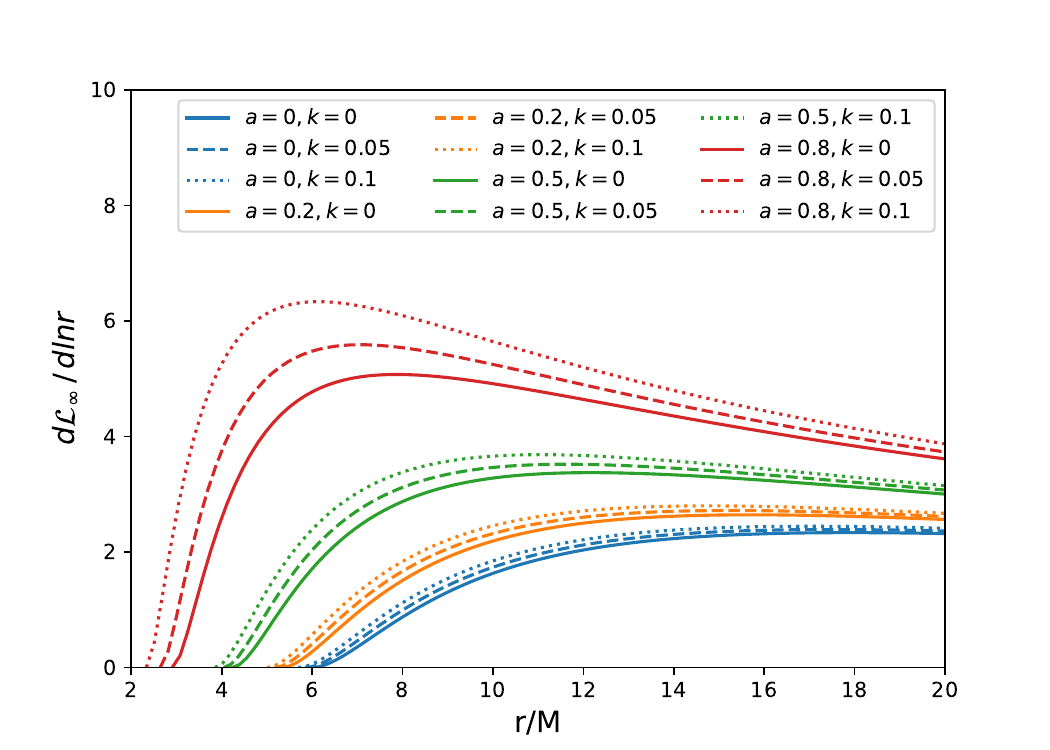}
\caption{Differential luminosity of thin accretion disk $d\mathcal{L}_{\infty}/d\ln r$ as a function of $r/M$ for different black hole spin $a$ and regular parameter $k$. The results of Kerr black hole ($k = 0$) are also provided as references. The values in this figure are multiplied by $10^2$.}
\label{d}
\end{figure}

Another interesting quantity we wish to study is the so-called differential luminosity of the thin accretion disk, which could be measured by observations. It takes the form as 
\begin{equation}
\frac{d \mathcal{L}_{\infty}}{d \ln r}=4 \pi r \sqrt{-g} E \mathcal{F}(r),
\end{equation}
where $\mathcal{L}_{\infty}$ represents the energy per unit of time reaching an observer at infinity. We have shown the differential luminosity as a function of radial distance with different black hole spin $a$ and regular parameter $k$ in Fig.~{\ref{d}}. The values in this figure are multiplied by $10^2$ for convenience.

For slowly rotating regular black holes, the differential luminosity for different $k$ values are nearly identical, suggesting a weak dependence on the regular parameter at low black hole spin. As the spin increases, the differential luminosity changes more significantly with the regular parameter $k$, particularly in the region near the black hole. The larger black hole spin $a$ and regular parameter $k$, the larger overall differential luminosity. The inner radii where differential luminosity approach to zero are also the corresponding $r_{\text{ISCO}}$. At larger radii, all curves converge, indicating that the effects of $a$ and $k$ diminish in the far region.

\section{Bolometric image of thin accretion disk around rotating regular black hole}\label{sec4}

\subsection{Ray-tracing method}

In order to obtain the black hole image, we will apply the ray-tracing method. By shooting light rays from an observer's image plane backward to the vicinity of the black hole, we could compute the resulting intensity in the image plane based on how the light rays intersect with the emission source (or the thin accretion disk) around the black holes. 

The conserved quantities of 
a photon $(\lambda, \eta)$ will determine its position on the image plane $(\alpha, \beta)$ \cite{ab}. Suppose an observe locate in $r_o$ relative to the black hole center and with inclination angle $\theta_o$, one could have the following relation \cite{ab}, 
\begin{equation}
\alpha=-\frac{\lambda}{\sin \theta_{o}}, \quad \beta = \pm_{o} \sqrt{\eta+a^{2} \cos ^{2} \theta_{o}-\lambda^{2} \cot ^{2} \theta_{o}},
\end{equation}
where $\pm_{o}$ is the sign of $\cos(\theta_{o})$, reversely
\begin{equation}\label{le}
\lambda=-\alpha \sin \theta_{o},\quad \eta=\left(\alpha^{2}-a^{2}\right) \cos ^{2} \theta_{0}+\beta^{2},
\end{equation}
The conserved quantities corresponding to photon shells or bounded orbits $(\tilde{\lambda}, \tilde{\eta})$ (see Eq.~\ref{cr1}-\ref{cr3}) will result in the critical curve $\mathcal{C} = \{\tilde{\alpha}, \tilde{\beta} \}$ in the image plane since they are not stable. The enclose region by this critical curve  in the image plane is usually denoted as the Black Hole Shadow. This critical curve will branch the behavior of light rays when we shoot backward from the image plane to the rotating regular black hole. For the light rays inside the critical curve, they will fall into black hole. While the light rays outside the critical curve are all scattered back to infinity, which means that they have a turning point in their trajectory. The radial turning point is the largest root of the radial potential $r_4$, and it is greater than the event horizon.

The relation Eq.~(\ref{le}) is not enough to determine the source coordinates of a photon. For a given black hole spin $a$, observer radial location $r_o$ and inclination angle $\theta_o$, our goal is to know the relation between the coordinates of the image plane $(\alpha, \beta)$ and the source coordinates $(r_s, \theta_s, \phi_s, t_s)$ where this photon emitted from. To do so, we need parameterize the photon trajectories by integrating the geodesic equations (\ref{pr}-\ref{pt}). In reference to \cite{lk1,lk2}, the differential geodesic equation in rotating regular black hole spacetime could also recast into an integral form,
\begin{align}
I_{r} & =G_{\theta}, \label{ig} \\
\Delta \phi \equiv \phi_{o}-\phi_{s} & =I_{\phi}+\lambda G_{\phi}, \label{dp}\\
\Delta t \equiv t_{o}-t_{s} & =I_{t}+a^{2} G_{t},\label{dt}
\end{align}
where the radial integrals
\begin{align}
I_{r} & =f_{r_{s}}^{r_{o}} \frac{\mathrm{~d} r}{ \pm_{r} \sqrt{\mathcal{R}(r)}},\label{ir}\\
I_{\phi}  &=f_{r_{s}}^{r_{o}} \frac{a(2 M r e^{-k/r}-a \lambda)}{ \pm_{r} \Delta(r) \sqrt{\mathcal{R}(r)}} \mathrm{d} r, \label{ip}\\
I_{t} & =f_{r_{s}}^{r_{o}} \frac{r^{2} \Delta(r)+2 M r e^{-k/r} \left(r^{2}+a^{2}-a \lambda\right)}{ \pm_{r} \Delta(r) \sqrt{\mathcal{R}(r)}} \mathrm{d} r, \label{it}  
\end{align}
and the angular integrals
\begin{align}
G_{\theta} & =f_{\theta_{s}}^{\theta_{o}} \frac{\mathrm{~d} \theta}{ \pm_{\theta} \sqrt{\Theta(\theta)}},\label{g8}\\
G_{\phi} &=f_{\theta_{s}}^{\theta_{o}} \frac{\csc ^{2} \theta}{ \pm_{\theta} \sqrt{\Theta(\theta)}} \mathrm{d} \theta, \label{gp}\\
G_{t} & =f_{\theta_{s}}^{\theta_{o}} \frac{\cos ^{2} \theta}{ \pm_{\theta} \sqrt{\Theta(\theta)}} \mathrm{d} \theta,\label{gt}
\end{align}
where the integral notation $f$ means the path integrals along the photon trajectory, and these path integrals are monotonically increasing along the trajectory. We could see that the angular integrals of rotating regular black hole are same to Kerr black hole \cite{lk1,lk2}.
However, the radial integrals are significantly different. For Kerr black hole, the radial integrals are analytical and revertible. However, it is not the case for rotating regular black hole, and we can only compute these radial integrals numerically. Anyway, by inverting these integrals, we are able to know the source coordinates given a photon with conservative quantities $(\lambda, \eta)$. 

Once we know the source coordinates, we can obtain the corresponding flux at this spacetime point according to the emission profile, and then adding photons accordingly to the image plane at position $(\alpha, \beta)$. Finally, we could obtain the black hole bolometric image. In this work, we consider a stationary axisymmetric thin accretion disk, so the emitting source only depends on the radius $r_s$. Every source ring with $r_s$ at the equatorial source plane (or the thin accretion disk) could have multiple images in the image plane, since they could be lensed multiple times by the rotating regular black hole. If a photon has cross the equatorial plane with $n$ times along its trajectory between the source and observer, then it will arrive at the image plane within a specific region where we called $n$th lensing band. For example, $n=0$ lensing band is just the direct image without any equatorial crossing in their trajectory. For the photons which have equatorial crossing once, it will located into the $n=1$ lensing band. The large $n$, the dimmer and narrower of the lensing bands, the closer to the critical curve \cite{lk1,lk2}. Therefore, we need to add up the photons in all the lensing bands to get a composite images of the black hole. We will only consider the lensing bands up to $n=2$ since it is almost coincide with the critical curve and much dimmer than the leading lensing bands. According to the radiative flux profile at the source coordinates, we can obtain the observed flux in the image plane according to the following relation,
\begin{equation}\label{oflux}
\mathcal{F}_{0}(\alpha, \beta)=\sum_{n=0}^{2}  \chi ^{4}\left(r_{\mathrm{s}}^{(n)}, \alpha, \beta\right) \mathcal{F}_{\mathrm{s}}\left(r_{\mathrm{s}}^{(n)}\right),
\end{equation}
where $\chi $ is the redshift factor given by
\begin{equation}
\chi = \frac{1}{u^t(1-\lambda \Omega)},
\end{equation}
$u^t$ is given by Eq.~(\ref{ut}). 

To summarize, the ray-tracing method has the following structure, and the steps are given by the corresponding equations discussed above,
\begin{align}
&\qquad \qquad \quad\, \text{(\ref{le})} \qquad \qquad \text{(\ref{g8},\,\ref{ig})}\nonumber\\
&\text{image}\, (\alpha, \beta)\; \rightarrow \; \text{rays} \,(\lambda,\eta)\; \rightarrow \text{integrals}\,(G_{\theta},I_r) \nonumber\\
&\text{(\ref{oflux})}\uparrow \qquad \qquad \qquad \qquad \qquad \qquad \qquad  \downarrow \text{(\ref{dp}-\ref{gt})}\qquad\nonumber\\
\; &\text{radiative flux profile}\, (\mathcal{F}_s) \;\longleftarrow \;
 \text{source}\, (r_s,\Delta \phi,\Delta t )\nonumber\\
 &\qquad \quad \qquad \qquad \qquad \qquad \;\text{(\ref{flux})} \nonumber
\end{align}

\subsection{Bolometric images of rotating regular black hole}

By employing the above ray-tracing method, we modified the well-known
$\textbf{aart}$\footnote{\url{https://github.com/iAART/aart}} code \cite{aart} to allow us to compute the observables numerically, and thus extend the code applicability to non-Kerr black holes. We investigate the lensing bands and bolometric images of rotating regular black holes as functions of black hole spin, the regular parameter, and the inclination angle. Furthermore, we perform a comparative analysis the results of rotating regular black hole with Kerr black hole, highlighting the differences induced by the presence of the regular parameter.

\begin{figure*}[htbp]
  \centering
    \includegraphics[scale = 0.48]{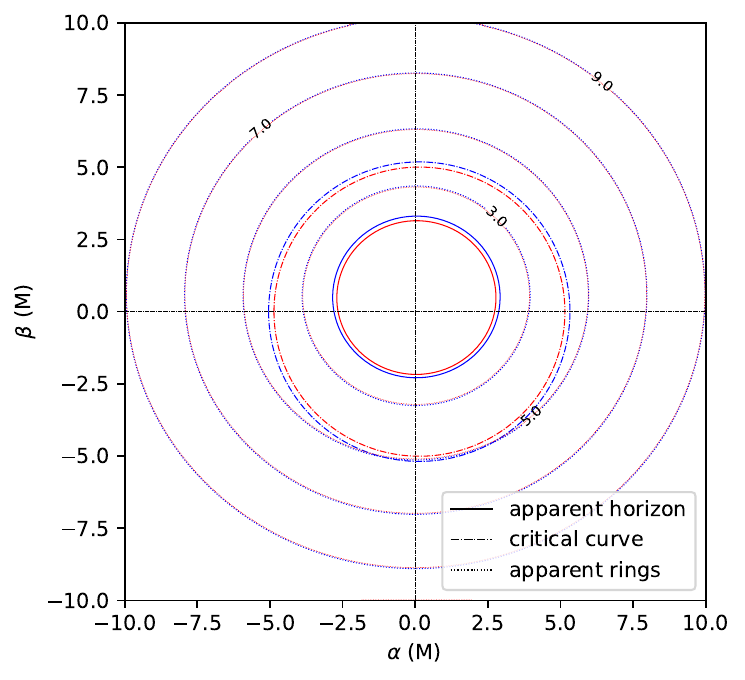}
    \includegraphics[scale = 0.48]{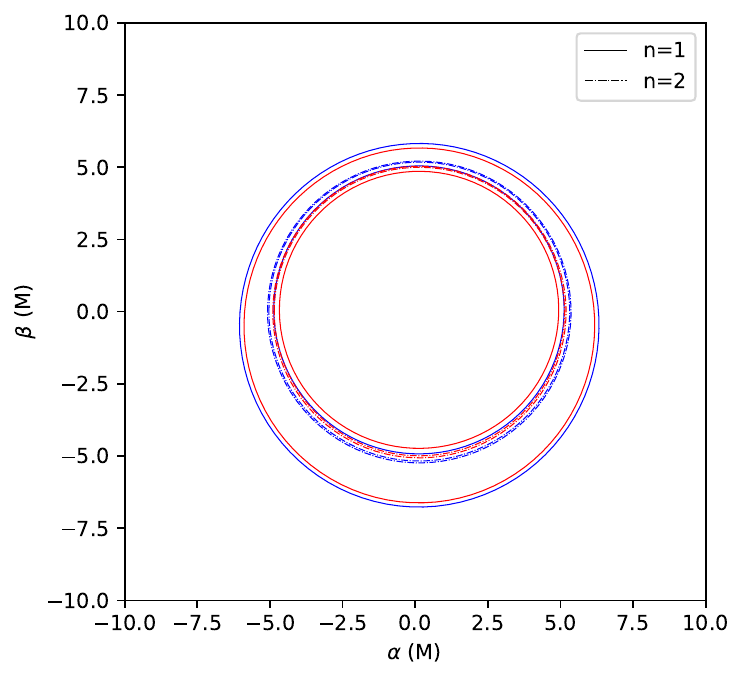}
    \includegraphics[scale = 0.48]{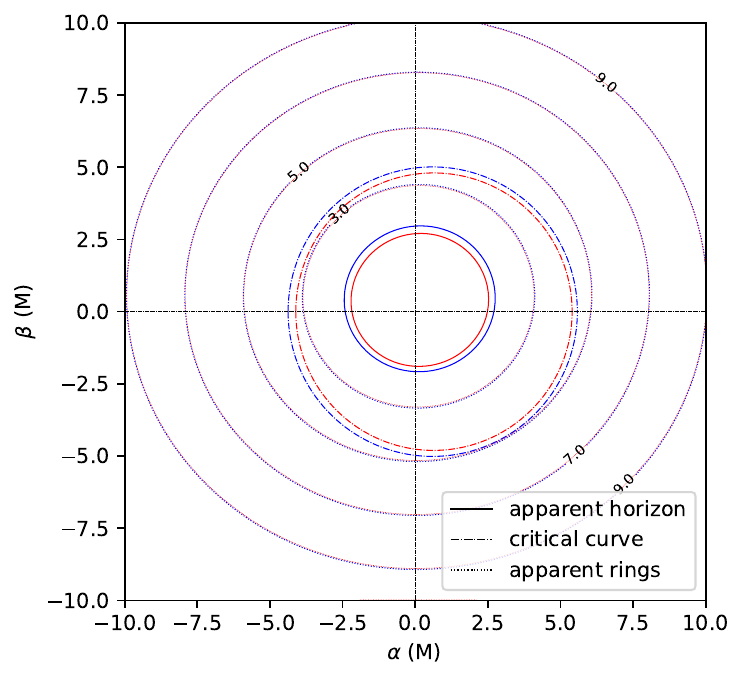}
    \includegraphics[scale = 0.48]{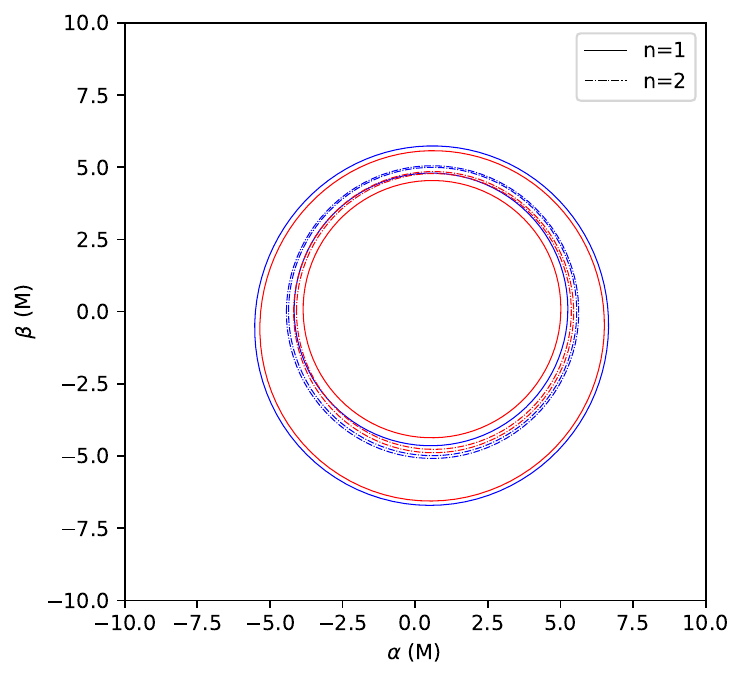}
    \includegraphics[scale = 0.48]{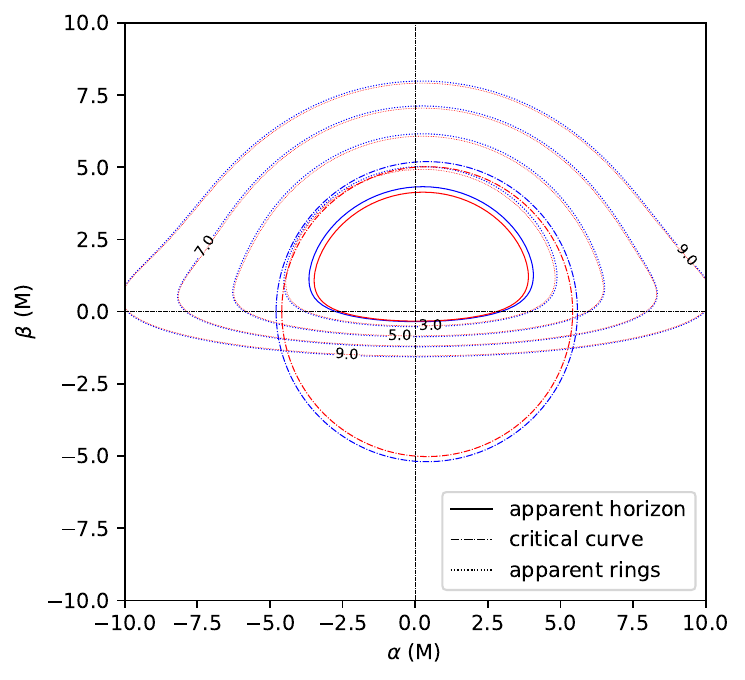}
    \includegraphics[scale = 0.48]{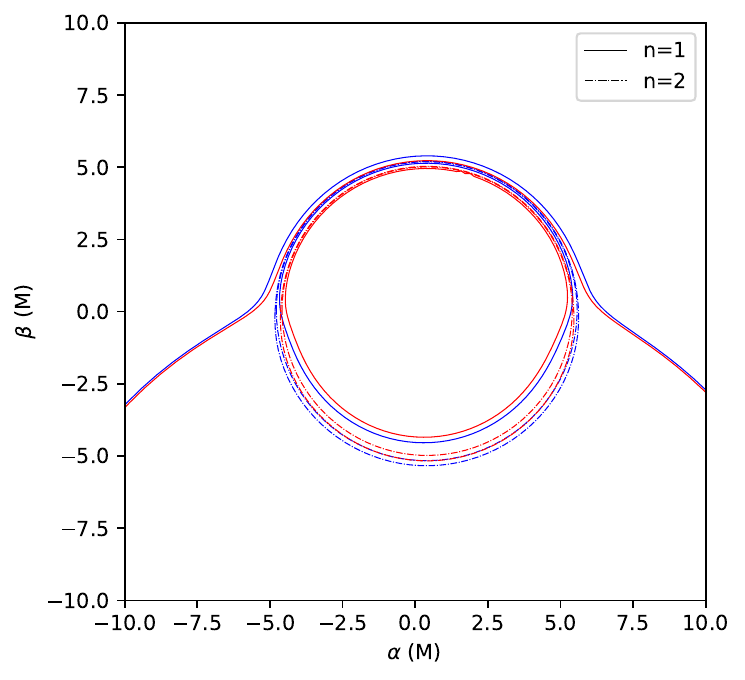}
    \includegraphics[scale = 0.48]{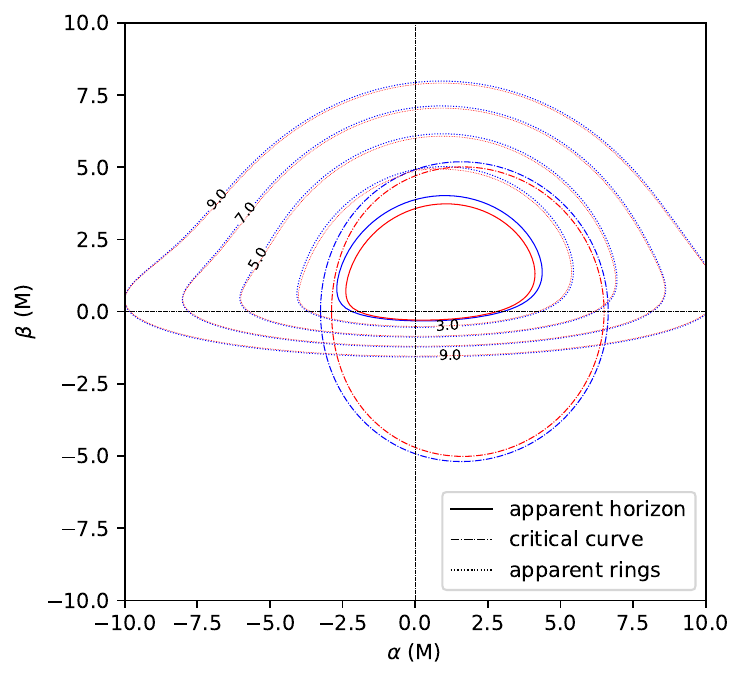}
    \includegraphics[scale = 0.48]{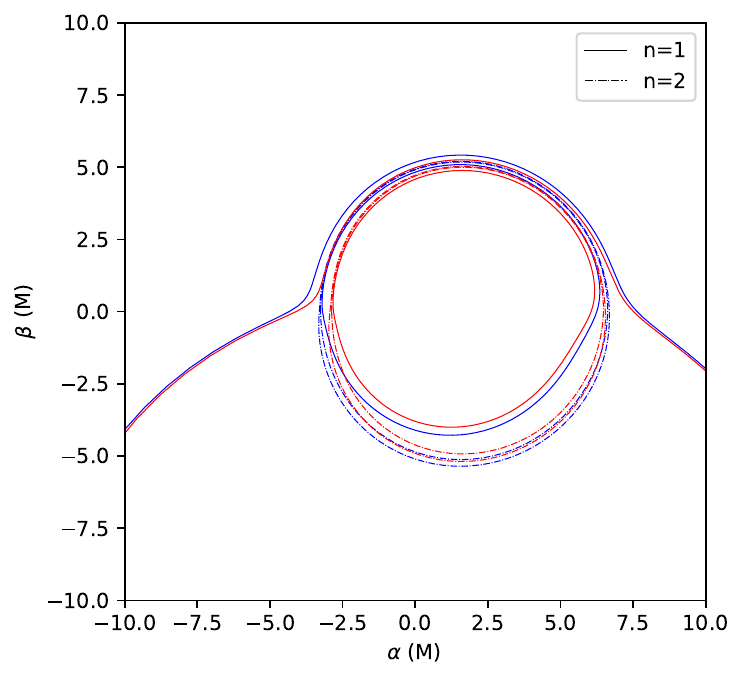}
\caption{Comparison of lensing bands for Kerr black holes (blue lines) and rotating regular black holes with k=0.1 (red lines). The first to fourth rows correspond to black hole spins and inclination angels with $(a=0.2, i=20^\circ), (a=0.8, i=20^\circ), (a=0.2, i=80^\circ), \text{and} (a=0.8, i=80^\circ)$, respectively. The left column shows direct image comparisons, including the apparent horizon, apparent source rings $(r_s =3,5,7,9)$, as well as the critical curve, while the right column compares the edges of n=1,2 lensing bands.}
\label{lb}
\end{figure*}

In Fig.~\ref{lb}, we present a comparative analysis of the lensing bands for Kerr black holes and rotating regular black holes with $k=0.1$. The comparison is made for different black hole spins ($a=0.2,\,0.8$) and inclination angles ($i=20^\circ,\,80^\circ$). The left column shows the comparison of the direct image (or $n=0$ lensing band), including the apparent horizon, apparent source rings ($r_s = 3, 5, 7, 9$), and the critical curve. The right column compares the edges of the lensing bands for $n=1$ and $n=2$. The blue lines represent the results for Kerr black holes, while the red lines correspond to the rotating regular black hole results.

From Fig.~\ref{lb}, we observe that for the direct image (or $n=0$ lensing band), the apparent horizons, apparent source rings, and critical curve are all smaller or compressed compared to the Kerr black hole, regardless of the black hole spin or inclination angle. These deviations arise due to the non-zero regular parameter $k$, and the effects of this parameter become more pronounced for larger black hole spins and higher inclination angles. The shrinking of the apparent source rings in the rotating regular black hole is particularly noticeable when the rings are closer to the horizon, as well as at larger inclination angles, while it is almost the same in the far region and at low inclination angle. Additionally, the apparent horizons, apparent source rings, and critical curves for Kerr and rotating regular black holes appear more circular at low inclination angles, becoming increasingly asymmetric at higher inclinations.

For the lensing bands $n=1$ and $n=2$, the differences between Kerr and rotating regular black holes are most evident at higher spins. Again, based on the edges of the lensing bands, the rotating regular black hole exhibits a slightly narrower and more compact structure compared to the Kerr results, reflecting the deviation introduced by the regular parameter $k=0.1$. The edges of the $n=1$ lensing band for the Kerr and rotating regular black holes nearly coincide as the lines move outward at higher inclination angles.

\begin{figure*}[htbp]
  \centering
    \includegraphics[scale = 0.55]{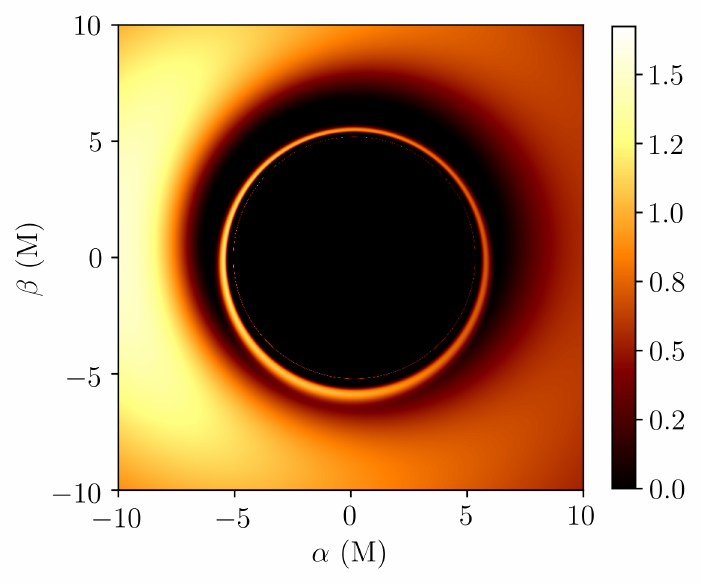}
    \includegraphics[scale = 0.55]{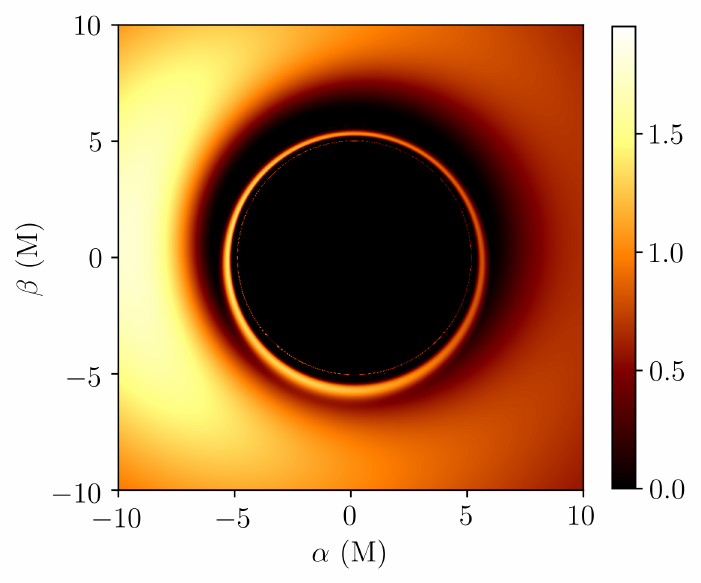}
    \includegraphics[scale = 0.55]{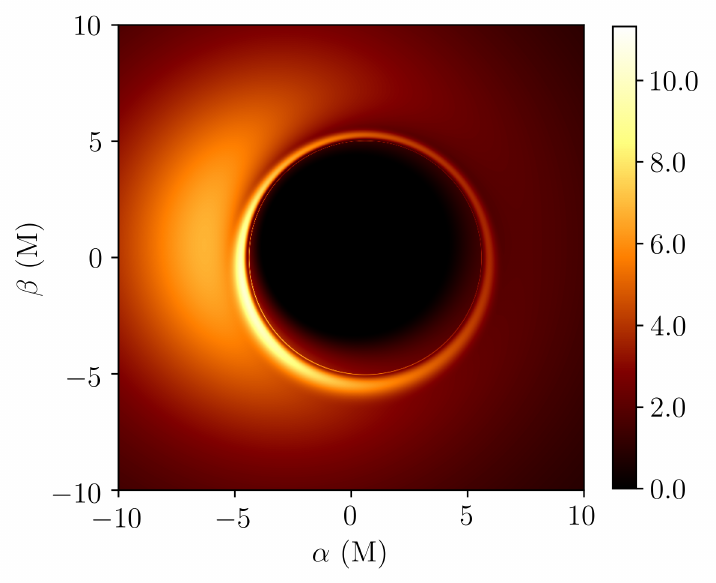}
    \includegraphics[scale = 0.55]{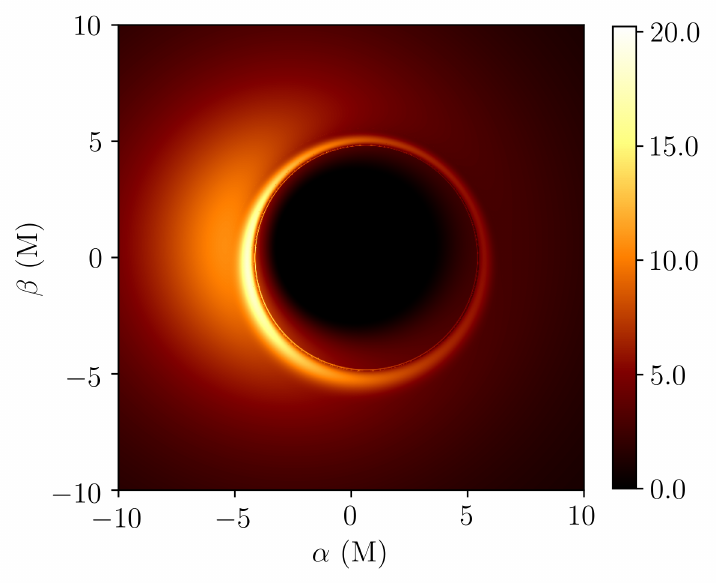}
    \includegraphics[scale = 0.55]{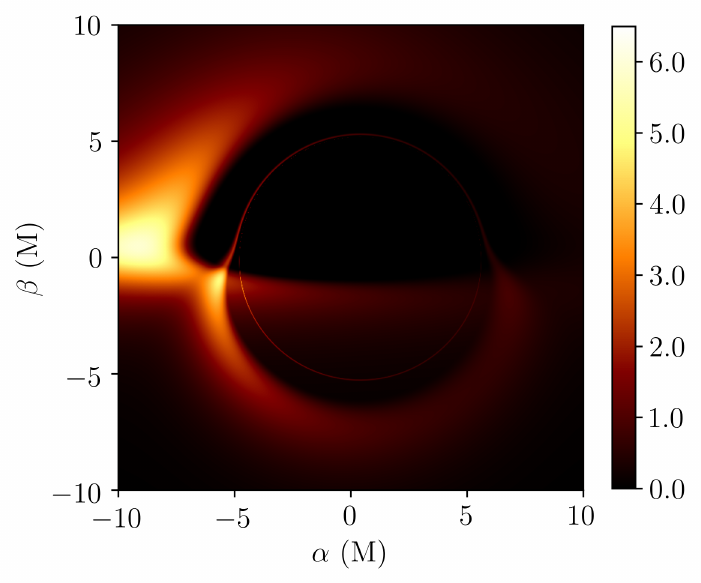}
    \includegraphics[scale = 0.55]{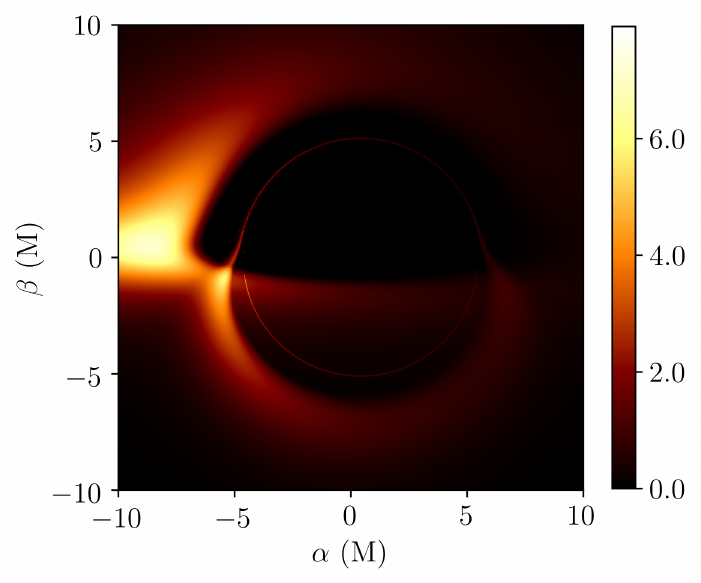}
    \includegraphics[scale = 0.55]{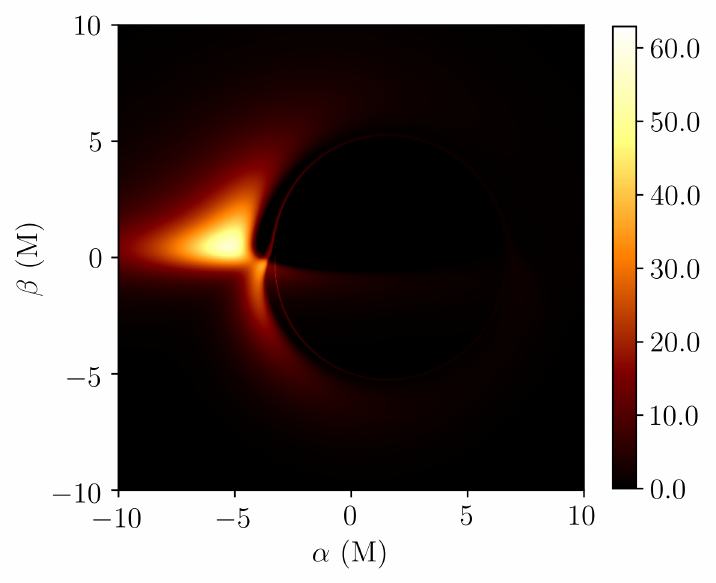}
    \includegraphics[scale = 0.55]{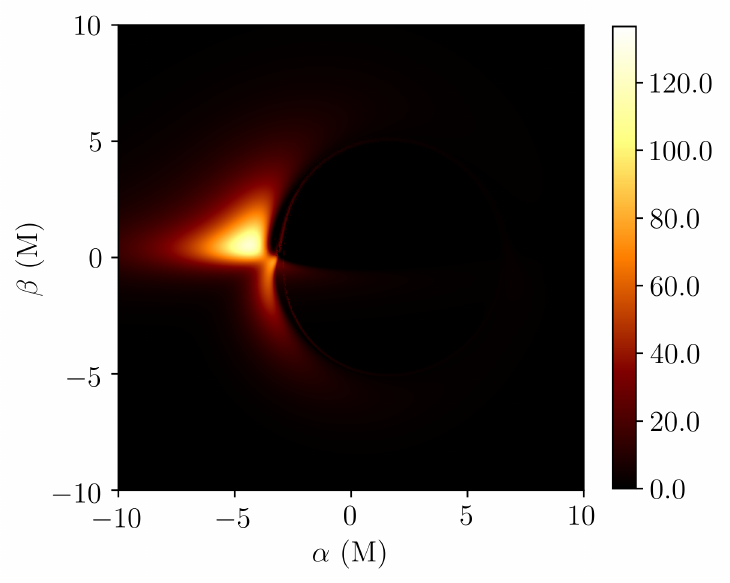}
\caption{Comparison of black hole images for Kerr black holes (left column) and regular black holes with $k=0.1$ (right column). The first to fourth rows correspond to parameters $(a=0.2, i=20^\circ), (a=0.8, i=20^\circ), (a=0.2, i=80^\circ)$,and $(a=0.8, i=80^\circ)$, respectively. The color bars in each subplot represent the brightness of the black hole image, with brighter regions corresponding to higher bolometric values. The source flux profile $\mathcal{F}_s(r)$ for generating the observed images are taken from Fig.~\ref{f} by making $\dot m =1$ and multiplying the values with a factor of $10^5$.}
\label{bhimage}
\end{figure*}

Next, we examine the bolometric image of the rotating regular black hole. In Fig.~\ref{bhimage}, we compare the bolometric images of the Kerr black hole and the rotating regular black hole with $k=0.1$, as a function of black hole spins ($a=0.2,\, 0.8$) and inclination angles ($i=20^\circ,\,80^\circ$). The radiative flux profile $\mathcal{F}_s(r)$ for generating the observed bolometric images are taken from Fig.~\ref{f}, by setting $\dot m =1$ and multiplying the values with a factor of $10^5$.

From Fig.~\ref{bhimage}, we could gain an intuitive understanding on the differences between the Kerr black hole and rotating regular black hole. As we discussed earlier, the shrinking effects in the rotating regular black hole are evident. Consequently, the structure of the rotating regular black hole becomes more concentrated compared to the Kerr black hole. Additionally, the rotating regular black holes display steeper and brighter brightness distributions than the Kerr black hole, indicating the influence of the modified geometry with $k=0.1$.

As already shown in Fig.~\ref{f}, the features of the thin accretion disk are also evident for different black hole spins and the regular parameter. Specifically, the inner edge of the accretion disk ($r_{ISCO}$) is smaller, and the radiative flux values are higher for larger black hole spins and regular parameters.

The rotating regular black hole images generally exhibit higher contrast and more pronounced asymmetries, while Kerr black hole images produce smoother intensity gradient. These differences highlight the potential of the rotating regular black hole to generate distinctive observational features that could distinguish it from the Kerr black hole.

\section{Conclusion and discussion}\label{sec5}




Rotating regular black holes, as a valuable exploration beyond GR, provide an intriguing model by resolving spacetime singularities. To test the physical implications of this idea, we investigated the observational features of thin accretion disks around this rotating regular black holes. Specifically, we analyzed key quantities of both time-like and null-like geodesics, illustrating how the regular parameter affects black hole geometrical properties such as horizons, the photon shell, and the ISCO. Based on these foundational results, we further examined the flux, temperature, and differential luminosity of the thin accretion disk as functions of the regular parameter and black hole spin. Additionally, using ray-tracing techniques, we explored the lensing behavior of the black hole and constructed realistic images of its thin accretion disk, which shows a significant deviation from the standard Kerr metric in GR by comparing their images.

The results show that the effects of the regular parameter on thin accretion disk observables, such as flux, temperature, and differential luminosity, become increasingly pronounced for rapidly rotating regular black holes. The non-zero regular parameter enhances the radiation efficiency of thin accretion disks, making these effects more detectable with observations. In the context of lensing, the regular parameter leads to a compression of lensing bands, a more compact black hole shadow, and a steeper brightness distributions, further highlighting the observational features of regular black holes.

Future research could aim to validate the existence of regular black hole through detailed observational data of accretion disk phenomena, particularly from instruments like the Event Horizon Telescope, may provide critical insights into the observational signatures predicted by our study. Beyond the thin accretion disk model, exploring other more complicate but realistic accretion disk models could also deepen our understanding of how spacetime regularity alters the physical processes around black holes. The study of rotating regular black holes will contribute to our understanding of black hole physics as well as the strong gravity regimes.

\section*{Acknowledgments}
Zhen Li would like to acknowledge the School of Science for supporting this study. Zhen Li is also financially supported by Start-up Funds for Doctoral Talents of Jiangsu University of Science and Technology. 
\\
\\


\begin{thebibliography}{0}%
\makeatletter
\providecommand \@ifxundefined [1]{%
 \@ifx{#1\undefined}
}%
\providecommand \@ifnum [1]{%
 \ifnum #1\expandafter \@firstoftwo
 \else \expandafter \@secondoftwo
 \fi
}%
\providecommand \@ifx [1]{%
 \ifx #1\expandafter \@firstoftwo
 \else \expandafter \@secondoftwo
 \fi
}%
\providecommand \natexlab [1]{#1}%
\providecommand \enquote  [1]{``#1''}%
\providecommand \bibnamefont  [1]{#1}%
\providecommand \bibfnamefont [1]{#1}%
\providecommand \citenamefont [1]{#1}%
\providecommand \href@noop [0]{\@secondoftwo}%
\providecommand \href [0]{\begingroup \@sanitize@url \@href}%
\providecommand \@href[1]{\@@startlink{#1}\@@href}%
\providecommand \@@href[1]{\endgroup#1\@@endlink}%
\providecommand \@sanitize@url [0]{\catcode `\\12\catcode `\$12\catcode `\&12\catcode `\#12\catcode `\^12\catcode `\_12\catcode `\%12\relax}%
\providecommand \@@startlink[1]{}%
\providecommand \@@endlink[0]{}%
\providecommand \url  [0]{\begingroup\@sanitize@url \@url }%
\providecommand \@url [1]{\endgroup\@href {#1}{\urlprefix }}%
\providecommand \urlprefix  [0]{URL }%
\providecommand \Eprint [0]{\href }%
\providecommand \doibase [0]{https://doi.org/}%
\providecommand \selectlanguage [0]{\@gobble}%
\providecommand \bibinfo  [0]{\@secondoftwo}%
\providecommand \bibfield  [0]{\@secondoftwo}%
\providecommand \translation [1]{[#1]}%
\providecommand \BibitemOpen [0]{}%
\providecommand \bibitemStop [0]{}%
\providecommand \bibitemNoStop [0]{.\EOS\space}%
\providecommand \EOS [0]{\spacefactor3000\relax}%
\providecommand \BibitemShut  [1]{\csname bibitem#1\endcsname}%
\let\auto@bib@innerbib\@empty
\end{thebibliography}%


\begin{thebibliography}{100}
\bibitem{gw1}
B. P. Abbott et al. (LIGO Scientific, Virgo), Phys. Rev. Lett. \textbf{116}, 061102 (2016).
\bibitem{gw2}
B. P. Abbott et al. (LIGO Scientific, Virgo), Phys. Rev.D. \textbf{100}, 104036 (2019).
\bibitem{gw3}
B. P. Abbott et al. (LIGO Scientific, Virgo), Phys. Rev. Lett. \textbf{119}, 161101 (2017).
\bibitem{bh1}
K. Akiyama et al. (Event Horizon Telescope), Astrophys.J. Lett. \textbf{875}, L1 (2019).
\bibitem{bh2}
K. Akiyama et al. (Event Horizon Telescope), Astrophys.J. Lett. \textbf{930}, L15 (2022).
\bibitem{sg1}
R. Penrose, Phys. Rev. Lett. \textbf{14}, 57 (1965).
\bibitem{sg2}
S. W. Hawking and R. Penrose, Proc. Roy. Soc. Lond. A. \textbf{314}, 529 (1970).
\bibitem{rr0}
R.~Torres, [arXiv:2208.12713 [gr-qc]].
\bibitem{rr1}
C.~Lan, H.~Yang, Y.~Guo and Y.~G.~Miao, Int. J. Theor. Phys. \textbf{62}, 202 (2023).
\bibitem{rr2}
C.~Bambi,[arXiv:2307.13249 [gr-qc]].
\bibitem{bar}
J. M. Bardeen, in Conference Proceedings of GR5, p. 174. Tbilisi University Press, 1968.
\bibitem{hay}
S. Hayward, Phys. Rev. Lett. \textbf{96}, 031103 (2006).
\bibitem{re1}
X. Li, Y. Ling, and Y.-G. Shen, Int. J. Mod. Phys. D. \textbf{22}, 1342016 (2013).
\bibitem{re2}
H. Culetu. Int. J. Theor. Phys. \textbf{54}, 2855 (2015).
\bibitem{re3}
K.~Pal, K.~Pal and T.~Sarkar, Gen. Rel. Grav. \textbf{55}, 121 (2023).
\bibitem{rre1}
S. G. Ghosh, Eur. Phys. J. C. \textbf{75}, 532 (2015)
\bibitem{rre2}
A. Simpson and M. Visser. JCAP.  \textbf{011}, 03 (2022).
\bibitem{rre3}
A. Simpson and M. Visser. Phys. Rev. D. \textbf{105}, 064065 (2022).
\bibitem{pre1}
M. Amir and S. G. Ghosh, Phys. Rev. D. \textbf{94}, 024054 (2016).
\bibitem{pre2}
R. Kumar and S. G. Ghosh, Class. Quant. Grav. \textbf{38}, 8 (2021).
\bibitem{pre3}
R. Kumar, A. Kumar and S. G. Ghosh, Astrophys. J. \textbf{896}, 89 (2020).
\bibitem{pre4}
F. Sarikulov, F. Atamurotov, A. Abdujabbarov, and B. Ahmedov, Eur. Phys. J. C. \textbf{82}, 771 (2022).
\bibitem{pre40}
Y.~Ling and M.~H.~Wu, Symmetry \textbf{14}, 2415 (2022). 
\bibitem{pre41}
M.~Guerrero, G.~J.~Olmo, D.~Rubiera-Garcia and D.~S\'aez-Chill\'on G\'omez, Phys. Rev. D \textbf{106}, 044070 (2022).
\bibitem{pre42}
S.~Vagnozzi, et al. Class. Quant. Grav. \textbf{40}, 165007 (2023).
\bibitem{pre5}
Z. Li, Phys. Rev. D. \textbf{107}, 044013 (2023).
\bibitem{pre6}
Z.~Li, X.~K.~Guo and F.~Yuan, Phys. Rev. D. \textbf{108}, 044067 (2023).
\bibitem{pre7}
L.~F.~D.~da Silva, F.~S.~N.~Lobo, G.~J.~Olmo and D.~Rubiera-Garcia, Phys. Rev. D. \textbf{108}, 084055 (2023).
\bibitem{pre8}
W.~Zeng, Y.~Ling, Q.~Q.~Jiang and G.~P.~Li, Phys. Rev. D. \textbf{108}, 104072 (2023). 
\bibitem{pre9}
D.~Zhang, H.~Gong, G.~Fu, J.~P.~Wu and Q.~Pan, Eur. Phys. J. C. \textbf{84}, 564 (2024).
\bibitem{pre10}
B.~Yang, G.~He, Y.~Xie and W.~Lin, Eur. Phys. J. C. \textbf{84}, 907 (2024).
\bibitem{pre11}
C.~A.~Benavides-Gallego, S.~Shashank and H.~Xu, [arXiv:2411.13897 [gr-qc]].
\bibitem{pre12}
M.~Calz\`a, D.~Pedrotti and S.~Vagnozzi,[arXiv:2409.02804 [gr-qc]].
\bibitem{ao1}
R. Edelson et al. Astrophys. J. \textbf{870}, 123 (2019).
\bibitem{ao2}
W. J. Guo et al. Astrophys. J. \textbf{929}, 19 (2022).
\bibitem{ao3}
Sathyaprakash R., et al. MNRAS. \textbf{511}, 5346 (2022).
\bibitem{ao4}
Y. Shen et al.  Astrophys. J, Supplement Series. \textbf{276}, 26 (2024).
\bibitem{lk1}
S. E. Gralla and A. Lupsasca. Phys. Rev. D. \textbf{101}, 044031 (2020).
\bibitem{lk2}
S. E. Gralla, A. Lupsasca and D. P. Marrone, Phys. Rev. D. \textbf{102}, 124004 (2020).
\bibitem{t1}
I. D. Novikov, K. S. Thorne, Black Holes (Les Astres Occlus), New
York, 343 (1973). 
\bibitem{t2}
D. N. Page, K. S. Thorne, Astrophys. J. \textbf{191}, 499 (1974). 
\bibitem{t3}
N. I. Shakura and R. A. Sunyaev. A\&A. \textbf{24}, 337 (1973).
\bibitem{ts1}
E.~Kurmanov, et al. Astrophys. J. \textbf{925}, 210 (2022). 
\bibitem{ts1}
K.~Boshkayev, et al. Eur. Phys. J. Plus. \textbf{139}, 273 (2024).
\bibitem{ts2}
K.~Boshkayev, et al. Eur. Phys. J. C. \textbf{84}, 230 (2024). 
\bibitem{ts3}
Y.~Ravanal, G.~G\'omez and N.~Cruz, Phys. Rev. D. \textbf{108}, 8 (2023).
\bibitem{ts4}
G.~Mustafa, et al. Eur. Phys. J. C. \textbf{84}, 690 (2024).
\bibitem{ts5}
S.~Patra, B.~R.~Majhi and S.~Das, JCAP. \textbf{060}, 01 (2024).
\bibitem{ts6}
L.~A.~S\'anchez, Eur. Phys. J. C. \textbf{84}, 635 (2024).
\bibitem{ts7}
Y.~H.~Jiang and T.~Wang, Phys. Rev. D. \textbf{110}, 103009 (2024).
\bibitem{ts8}
O.~Layeghi, J.~Ghanbari and M.~Moeen Moghaddas, [arXiv:2409.02955 [gr-qc]].
\bibitem{ts9}
H.~B.~Zheng, M.~Q.~Wu, G.~P.~Li and Q.~Q.~Jiang,[arXiv:2411.10315 [gr-qc]].
\bibitem{ts10}
Y.~Hou, Z.~Zhang, H.~Yan, M.~Guo and B.~Chen, Phys. Rev. D. \textbf{106}, 064058 (2022).
\bibitem{ts11}
J.~Peng, M.~Guo and X.~H.~Feng, Chin. Phys. C. \textbf{45}, 085103 (2021). 
\bibitem{ts12}
Z.~Zhang, Y.~Hou, M.~Guo and B.~Chen, JCAP. \textbf{05}, 032 (2024).
\bibitem{geo}
J.~M.~Bardeen, W.~H.~Press and S.~A.~Teukolsky, Astrophys. J. \textbf{178}, 347 (1972).
\bibitem{ab}
C. T. Cunningham and J. M. Bardeen,  Astrophys. J. \textbf{183}, 237 (1973).
\bibitem{che}
H.~Chen, X.~Y.~Chew and W.~Fan, [arXiv:2411.00565 [gr-qc]].
\bibitem{aart}
A. Cardenas-Avendano, A. Lupsasca,  H. Zhu, Phys. Rev. D. \textbf{107}, 043030, (2023). arXiv:2211.07469.






























\end{thebibliography}
\end{document}